\documentclass[sigconf,nonacm,natbib=true]{acmart}
\AtBeginDocument{%
  }

\usepackage{csquotes}
\usepackage{mdframed}
\usepackage{enumitem}
\usepackage{tabularx}
\usepackage[htt]{hyphenat}
\usepackage{algorithm}
\usepackage{algorithmic}
\usepackage{amsmath} 
\usepackage{array}
\usepackage{float}
\usepackage{graphicx}
\usepackage{pbalance}
\usepackage{multirow}

\newcommand{\hflogo}{\raisebox{-0.18\height}{%
  \includegraphics[height=0.9em]{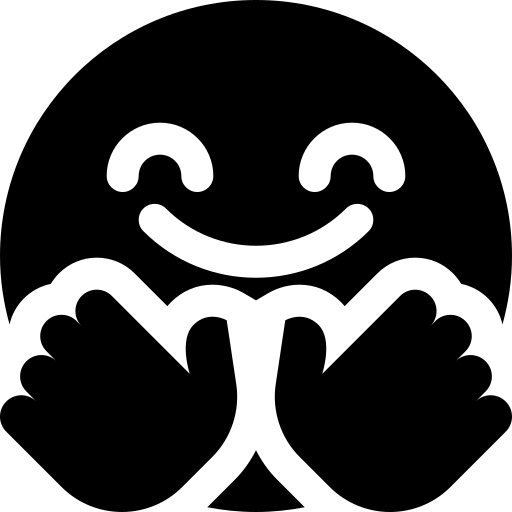}}}

\newmdenv[
  backgroundcolor=gray!8,
  linecolor=gray!50,
  roundcorner=4pt,
  skipabove=6pt,
  skipbelow=6pt,
  innerleftmargin=8pt,
  innerrightmargin=8pt,
  innertopmargin=6pt,
  innerbottommargin=6pt
]{prompt}

\begin{document}

\title{AgentSim: A Platform for Verifiable Agent-Trace Simulation}

\author{Saber Zerhoudi}
\orcid{0000-0003-2259-0462}
\affiliation{%
  \institution{University of Passau}
  \city{Passau}
  \country{Germany}
}
\email{saber.zerhoudi@uni-passau.de}

\author{Michael Granitzer}
\orcid{0000-0003-3566-5507}
\affiliation{%
  \institution{University of Passau}
  \city{Passau}
  \country{Germany}
}
\affiliation{%
  \institution{Interdisciplinary Transformation University Austria}
  \city{Linz}
  \country{Austria}
}
\email{michael.granitzer@uni-passau.de}

\author{Jelena Mitrovi\'{c}}
\orcid{0000-0003-3220-8749}
\affiliation{%
  \institution{University of Passau}
  \city{Passau}
  \country{Germany}
}
\email{jelena.mitrovic@uni-passau.de}

% \renewcommand{\shortauthors}{S. Zerhoudi et al.}

% --- ACM-like first-page rights styling ---
\makeatletter
% Slightly tighten footnote area spacing (like acmart)
\setlength{\skip\footins}{9pt plus 2pt minus 1pt}

% Font & layout for the rights block (≈ 8pt on 9.5pt leading)
\newcommand{\acmrightssize}{\fontsize{8}{9.5}\selectfont}

\setlength{\emergencystretch}{1.5em} % try 1em–3em

% keep this
\settopmatter{printacmref=false}

% Unnumbered, ACM-styled first-page footnote helper
\newcommand{\firstpagerights}[1]{%
  \begingroup
    \renewcommand\thefootnote{}%
    \footnotetext{%
      \acmrightssize
      \raggedright
      \setlength{\parskip}{0pt}%
      \setlength{\parindent}{0pt}%
      #1%
    }%
    \addtocounter{footnote}{0}%
  \endgroup
}
\makeatother

\begin{abstract}
Training trustworthy agentic LLMs requires data that shows the grounded reasoning process, not just the final answer. Existing datasets fall short: question-answering data is outcome-only, chain-of-thought data is not tied to specific documents, and web-agent datasets track interface actions rather than the core retrieval and synthesis steps of a RAG workflow. We introduce AgentSim, an open-source platform for simulating RAG agents. It generates verifiable, stepwise traces of agent reasoning over any document collection. AgentSim uses a policy to ensure the agent widely explores the document set. It combines a multi-model validation pipeline with an active human-in-the-loop process. This approach focuses human effort on difficult steps where models disagree. Using AgentSim, we construct and release the Agent-Trace Corpus (ATC), a large collection of grounded reasoning trajectories spanning three established IR benchmarks. We make three contributions: (1) the AgentSim platform~\footnote{\label{fn:platform}Platform: \url{https://agentsim.searchsim.org}} with two mechanisms, Corpus-Aware Seeding and Active Validation, that improve trace diversity and quality; (2) the Agent-Trace Corpus (ATC)~\footnote{\label{fn:atc}\hflogo~ATC Dataset: \url{https://huggingface.co/datasets/searchsim/agentsim-atc}}, over 103,000 verifiable reasoning steps spanning three IR benchmarks, with 100\% grounding rate on substantive answers; and (3) a comparative behavioral analysis revealing systematic differences in how state-of-the-art models approach information seeking~\footnote{\label{fn:toolkit}Repository: \url{https://github.com/searchsim-org/sigir26-agentsim}}. Platform, toolkit, and corpus are publicly available.
\end{abstract}




\begin{teaserfigure}
  \begin{center}
\includegraphics[width=.86\textwidth]{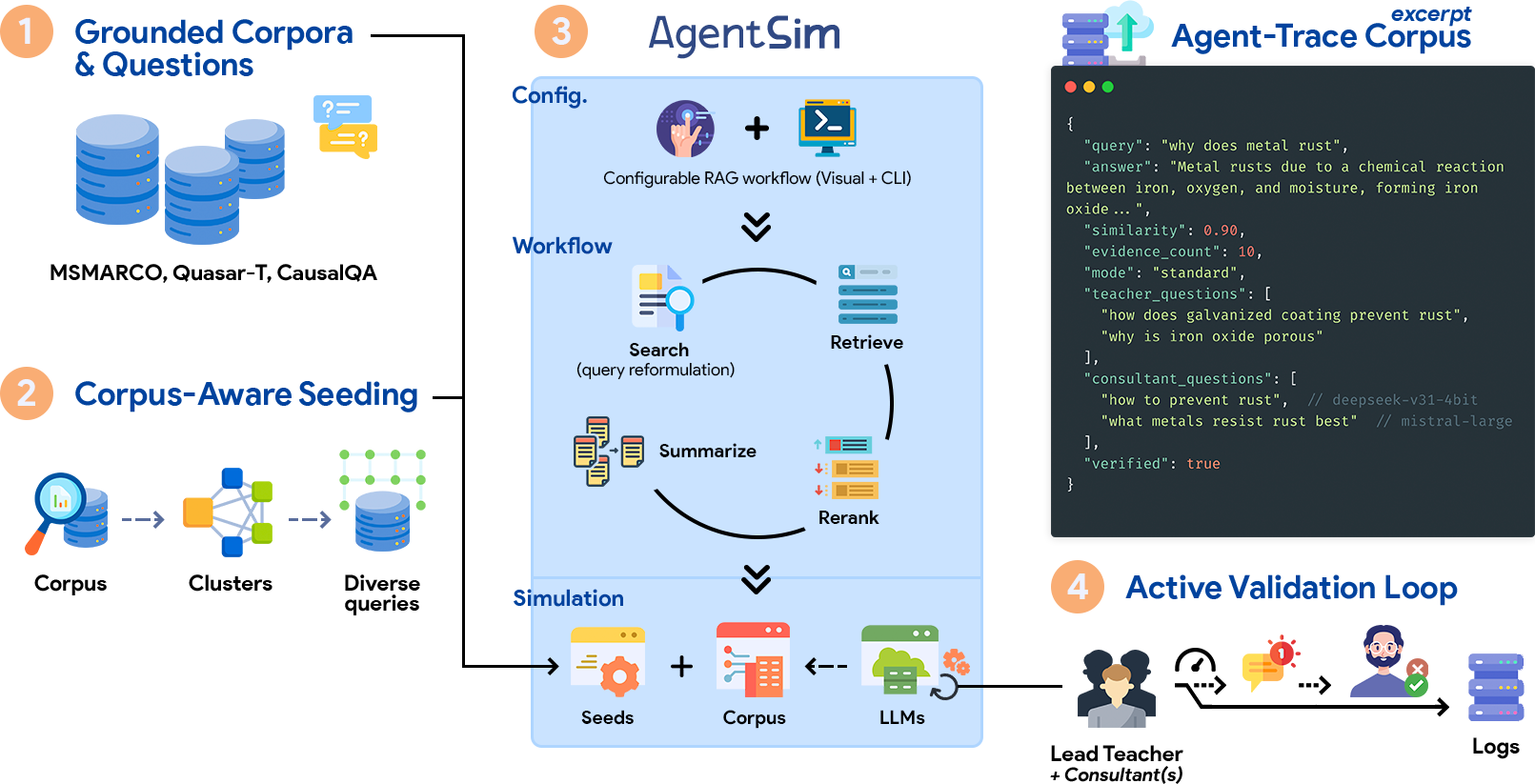}
\vspace{-3mm}
  \caption{Overview of the AgentSim Workflow and Agent-Trace Corpus.}
  \label{fig_simiir3}
  \vspace{4mm}
\end{center}
\end{teaserfigure}

\maketitle
\enlargethispage{2\baselineskip}
\firstpagerights{%
  © ACM, 2026. This is the author's version of the work.\\
  The definitive version was published in:
  \emph{Proceedings of the 49th International ACM SIGIR Conference on Research and Development in Information Retrieval (SIGIR '26), July 20--24, 2026, Melbourne, VIC, Australia}.\\
  DOI: \url{https://doi.org/10.1145/3805712.3808577}
}

\section{Introduction}
Large language models are evolving into autonomous agents capable of multi-step reasoning and tool use. This shift holds promise for information retrieval, yet it introduces a challenge: the internal reasoning of these agents remains hidden. When an agent fails to find relevant information or produces an incorrect synthesis, practitioners cannot see \emph{why}. This opacity makes agentic systems hard to trust, debug, and improve.

A path toward more auditable agents lies in training them on data that captures the full reasoning \emph{process}, not just the final \emph{outcome}. This is especially true for Retrieval-Augmented Generation (RAG)~\cite{Lewis:2020:NeurIPS}, where retrieval-driven user-centric agents are emerging~\cite{Zerhoudi:2024:PersonaRAG}. We identify what we call the \textbf{cognitive RAG gap}~\cite{Zerhoudi:2026:BeyondClick}: existing datasets do not capture the intermediate steps (query formulation, document evaluation, information synthesis, and strategy adaptation) that make up effective information seeking.

We can characterize this gap by examining three categories of existing resources. \emph{Question-answering datasets} such as MS~MARCO~\cite{Bajaj:2018:arXiv} and Natural Questions~\cite{Kwiatkowski:2019:TACL} provide query-answer pairs but omit the retrieval process entirely; they are outcome-only. \emph{Reasoning-trace datasets} like OpenThoughts~\cite{Guha:2025:arXiv} and PRM800K~\cite{Lightman:2023:arXiv} capture step-by-step reasoning for abstract tasks such as mathematics and coding, yet these traces are not grounded in external, verifiable documents. \emph{Web-agent benchmarks} including WebArena~\cite{Zhou:2024:arXiv} and Mind2Web~\cite{Deng:2024:NeurIPS} focus on interface navigation rather than the cognitive workflow of retrieval and synthesis.

This paper introduces \textbf{AgentSim}, an open-source platform designed to bridge this gap (Figure~\ref{fig_simiir3}). AgentSim enables the design, simulation, and recording of grounded agent traces over document collections. The system comprises a visual web interface for prototyping and a command-line toolkit for scalable data generation. Two mechanisms differentiate AgentSim from prior tools. First, \emph{Corpus-Aware Seeding} employs embedding-based clustering and novelty filtering to ensure simulations explore the full breadth of a target corpus, avoiding the redundancy of random sampling. Second, an \emph{Active Validation} loop employs multi-model disagreement to automatically flag ambiguous steps for human review, focusing annotation effort where it matters most.

Using AgentSim, we construct and publicly release the \textbf{Agent-Trace Corpus (ATC)}\textsuperscript{\ref{fn:atc}} , a collection of over 103,000 verifiable reasoning steps spanning the MS~MARCO, Quasar-T, and CausalQA~\cite{Bondarenko:2022:ICCL} corpora. We use this corpus to conduct a comparative analysis of three state-of-the-art models, identifying variations in exploration strategy, query reformulation behavior, and synthesis under uncertainty. The results support the utility of the platform as a research tool and a training source for RAG agents.

\section{Related Work}
Our work sits at the intersection of reasoning-trace datasets, RAG evaluation, agent simulation platforms, and synthetic data generation. We position AgentSim relative to each of these areas.

\textbf{Reasoning-Trace Datasets.}
Recent research captures explicit reasoning processes to improve model training. Datasets such as OpenThoughts~\cite{Guha:2025:arXiv} and PRM800K~\cite{Lightman:2023:arXiv} provide step-by-step traces for mathematics and code generation, training models to ``show their work''~\cite{Wei:2023:arXiv,Abdin:2025:arXiv}. However, such traces are self-contained: they capture internal reasoning but are not tied to external, verifiable knowledge sources. Our work differs in that every reasoning step in AgentSim links to concrete retrieval actions and specific documents, allowing verification against an external corpus.

\textbf{RAG Evaluation Benchmarks.}
The evaluation of retrieval-augmented systems relies on established collections such as MSMARCO~\cite{Bajaj:2018:arXiv}, Natural Questions~\cite{Kwiatkowski:2019:TACL}, and domain-specific resources like CausalQA~\cite{Bondarenko:2022:ICCL}. These benchmarks provide query-answer pairs for measuring system accuracy~\cite{Teixeira:2024:arXiv,Yu:2025:Springer}. Our platform does not replace these resources but rather builds upon them: we use their queries as starting points and their document collections as the environment for simulation. The contribution is transforming static query-answer pairs into dynamic traces that reveal the full information-seeking process, including failed retrievals, query reformulations, and iterative refinement.

\textbf{Agent Simulation Platforms.}
Interactive benchmarks such as WebArena~\cite{Zhou:2024:arXiv}, Mind2Web~\cite{Deng:2024:NeurIPS}, and ALFWorld~\cite{Shridhar:2021:arXiv} evaluate agents through interface-level interactions (clicking, navigating, form-filling). A parallel IR line produces user-simulation toolkits for interaction logs and behavioural data~\cite{Azzopardi:2024:SimIIR3,Balog:2025:UserSimToolkit}, but targets session-level user behaviour, not the cognitive workflow of an agentic retriever. AgentSim instead simulates and records the \emph{cognitive process} behind RAG tasks---how agents formulate queries, evaluate documents, synthesize information, and adapt strategies---a focus that distinguishes it from both world- and user-simulation tools.

\textbf{Synthetic Data Quality.}
Using LLM-generated data for training raises concerns about quality and diversity~\cite{Shumailov:2024:Nature}. Common mitigations include majority voting~\cite{Wang:2023:ICLR} and consistency filtering. Our Active Validation loop goes beyond passive voting by introducing targeted human review: rather than applying uniform quality checks, we use model disagreement (measured via a Divergence Score) to identify steps where automated validation falls short. This active approach focuses human annotation on ambiguous cases, improving cost-effectiveness while maintaining data quality~\cite{Wu:2024:arXiv,Pleias:2025:article}.

\section{System Architecture and Components}
AgentSim facilitates flexible and reproducible data generation. The platform combines a visual front-end for workflow design with a command-line toolkit for scalable trace creation. To improve synthetic data quality and coverage, the system employs two mechanisms: Corpus-Aware Seeding and Active Validation.

The platform's design follows recent findings that high-quality reasoning traces are data-efficient for training capable models~\cite{Pleias:2025:article}. Rather than maximizing trace volume, AgentSim prioritizes trace diversity and verifiability. The modular architecture lets researchers combine tools for retrieval, generation, and verification, supporting both behavioral analysis and systematic data creation.

\subsection{Platform and Toolkit}

The \emph{Platform}\textsuperscript{\ref{fn:platform}} provides a web-based visual interface for designing and testing agent workflows. Users construct pipelines by connecting modular components (search, rerank, summarize, synthesize) and can step through simulations interactively to prototype and debug agent behavior. A demonstration walkthrough is available in the repository documentation, guiding users through creating RAG workflows, configuring models, and inspecting generated traces.

The \emph{Toolkit}\textsuperscript{\ref{fn:toolkit}} is a command-line interface for large-scale data generation. It executes simulations from YAML configuration files specifying the agent's tools, models, corpus connection, and seeding parameters. Core commands include \texttt{\small agentsim simulate} for trace generation, \texttt{\small agentsim seed-select} for corpus-aware seed creation, and \texttt{\small agentsim validate} for configuration verification. These components form a unified system for repeatable data generation.

\subsection{Corpus-Aware Seeding Policy}
\label{seeding}

Random sampling of simulation starting points is inefficient: it tends to favor popular topics and produces redundant traces that waste computational resources. Our platform implements a \textbf{Corpus-Aware Seeding} policy to ensure systematic coverage of the target document collection while maximizing trace novelty.

Algorithm~\ref{alg:seeding} formalises the policy: given candidate queries $Q$, corpus $C$, and budget $B$, it produces a seed set $S$ that balances topical coverage with document novelty by jointly optimising two dimensions---query semantic diversity (different topics) and retrieval novelty (different regions of the document space). The encoder choice matters: retrieval behaviour varies substantially across encoder families on identical corpora~\cite{Caspari:2024:BeyondBenchmarks}, so we expose it as a hyperparameter (Table~\ref{tab:hyperparams}) and report results under \texttt{\small all-MiniLM-L6-v2}.

\begin{algorithm}[H]
\caption{Corpus-Aware Seed Selection}
\label{alg:seeding}
\begin{algorithmic}[1]
\small
\REQUIRE Candidate queries $Q$, corpus $C$, budget $B$
\ENSURE Diverse seed set $S$
\STATE $E \leftarrow \text{Embed}(Q)$ using SentenceTransformer
\STATE $\{C_1, \ldots, C_K\} \leftarrow \text{KMeans}(E, K)$
\STATE $\text{coverage}[c] \leftarrow 0$ for all clusters $c$
\STATE $S \leftarrow \emptyset$; $\text{seen\_docs} \leftarrow \emptyset$
\WHILE{$|S| < B$}
    \STATE $c^* \leftarrow \arg\min_c \text{coverage}[c]$
    \STATE $Q_c \leftarrow \{q \in Q : \text{cluster}(q) = c^*\}$
    \FOR{$q \in Q_c$}
        \STATE $D_q \leftarrow \text{Retrieve}(q, C)$
        \STATE $\text{novelty}(q) \leftarrow |D_q \setminus \text{seen\_docs}| / |D_q|$
    \ENDFOR
    \STATE Filter $Q_c$ to queries with $\text{novelty}(q) > \tau$
    \STATE $q^* \leftarrow \text{MMR}(Q_c, S, \lambda)$
    \STATE $S \leftarrow S \cup \{q^*\}$
    \STATE $\text{seen\_docs} \leftarrow \text{seen\_docs} \cup D_{q^*}$
    \STATE $\text{coverage}[c^*] \leftarrow \text{coverage}[c^*] + 1$
\ENDWHILE
\RETURN $S$
\end{algorithmic}
\end{algorithm}

Table~\ref{tab:hyperparams} lists the key hyperparameters. The policy outputs a \texttt{\small seeds.jsonl} file that enables transparent coverage measurement and supports controlled experiments where different agent configurations are compared over identical starting conditions.

\begin{table}[tbp]
\caption{Hyperparameters for Corpus-Aware Seeding.}
\label{tab:hyperparams}
\centering
\small
\setlength{\tabcolsep}{3pt}
\renewcommand{\arraystretch}{1.05}

\begin{tabularx}{\linewidth}{@{}p{0.3\linewidth}cX@{}}
\toprule
\textbf{Parameter} & \textbf{Value} & \textbf{Description} \\
\midrule
$K$ (clusters)     & 50    & Number of topic clusters \\
$\tau$ (novelty)   & 0.4   & Min.\ fraction of new documents \\
$\lambda$ (MMR)    & 0.7   & Diversity--relevance trade-off \\
Embedding          & MiniLM & \texttt{\small all-MiniLM-L6-v2} model \\
\bottomrule
\end{tabularx}

% \vspace{-2em}
\end{table}

\subsection{Active Validation and Refinement Loop}
\label{validation_loop}
Generating high-quality traces at scale requires balancing automation with human oversight. Our system implements an \textbf{Active Validation} loop that uses model disagreement to focus human review precisely where it is most needed.

The mechanism operates through an Analyst--Critic--Judge pipeline. For each simulation step, an Analyst model proposes an action (e.g., a query reformulation or synthesis). Two Critic models independently evaluate and potentially revise this proposal. A Judge component then computes a \textbf{Divergence Score} $\text{DS}(s) = 1 - {\max_{a \in A} |\{m : m \text{ chose } a\}|}\,/\,{|M|}$, where $A$ is the set of distinct actions proposed and $M$ is the set of participating models. A step is flagged for human review when $\text{DS}(s) > \theta$ (we use $\theta = 0.4$). For each flagged step, reviewers see the competing model outputs and choose to \emph{Promote} the best one, \emph{Revise} it, or \emph{Discard} the trajectory, following a rubric on query--document relevance, evidence sufficiency, and synthesis faithfulness. Inter-reviewer agreement is tracked via 10\% double-annotation, with disagreements adjudicated.

The loop has two independent triggers. The \emph{Divergence Score} routes ambiguous steps to human review. A separate, automated route uses a per-step \emph{grounding-confidence score} that measures how well the proposed answer is supported by its cited evidence: when this score is low ($<0.3$), the step is sent for automatic re-retrieval. This split keeps human attention on cases where models genuinely disagree, while still repairing low-confidence steps at scale~\cite{Wu:2024:arXiv,Pleias:2025:article}. Across 5,320 ATC steps, 73.4\% had low grounding confidence and triggered automatic re-retrieval; 99.8\% of those subsequently received external verification; and 79.3\% of seed queries showed improved grounding after iterative validation. 

\section{The Agent-Trace Corpus (ATC)}
Using AgentSim, we construct and release the \emph{Agent-Trace Corpus}, a large collection of grounded reasoning traces designed for training and analyzing RAG agents. This section describes the corpus statistics, data formats, and intended use cases.

\subsection{Corpus Statistics}
The ATC comprises traces generated across three established IR benchmarks, using three state-of-the-art language models. Table~\ref{tab:corpus_stats} summarizes the key statistics. The corpus contains 103,567 reasoning trace steps, each connecting an agent's internal deliberation to concrete retrieval actions and specific documents. From these traces, we extract 20,548 supervised training pairs suitable for fine-tuning, along with nearly 200,000 unique retrieved documents.

\begin{table}[tbp]
\caption{Agent-Trace Corpus (ATC) statistics.}
\label{tab:corpus_stats}
\centering
\small
\begin{tabular}{lr}
\toprule
\textbf{Statistic} & \textbf{Value} \\
\midrule
Total trace steps & 103,567 \\
Supervised training pairs & 20,548 \\
Unique documents retrieved & 199,968 \\
Total queries generated & 26,176 \\
\midrule
Source corpora & MS~MARCO, Quasar-T, CausalQA \\
Analyst models & \texttt{\footnotesize gpt-4o}, \texttt{\footnotesize mistral-large}, and \texttt{\footnotesize deepseek-v3}\\
\bottomrule
\end{tabular}

% \vspace{-2.1em} 
\end{table}

The three source corpora provide complementary coverage: MS~MARCO contributes web passage retrieval tasks, Quasar-T provides factual question answering, and CausalQA offers complex causal reasoning challenges. This diversity ensures the corpus exercises different aspects of agent reasoning.

\subsection{Data Formats}
The ATC is distributed in three formats, serving distinct purposes:

\emph{Traces} (\texttt{\small traces/*.jsonl.gz}) contain the full reasoning sequence for each simulation: thought $\rightarrow$ action $\rightarrow$ observation chains with complete LLM inputs and outputs. Each trace records the agent's internal deliberation (``I need to find information about X''), the action taken (``search for Y''), and the retrieved results. These are suitable for behavioral analysis and process-level supervision.

\emph{Trajectories} (\texttt{\small trajectories/*.jsonl.gz}) provide high-level action sequences abstracted from the raw LLM outputs. Each trajectory lists the sequence of tools called (search, rerank, summarize) along with their inputs and outputs, without the full prompt text. These compact representations are efficient for training action-prediction models and studying exploration patterns.

\emph{Supervised pairs} (\texttt{\small supervised/*.jsonl.gz}) extract query--document--answer triples with multi-hop reasoning chains. Each record contains a question, the documents retrieved to answer it, and the final synthesized answer. These are formatted for direct use in supervised fine-tuning pipelines.

\subsection{Intended Use Cases}
The corpus supports multiple training paradigms. For \emph{chain-of-thought fine-tuning}, the traces provide explicit reasoning grounded in retrieved documents. In \emph{imitation learning}, the trajectories demonstrate effective information-seeking behavior. For \emph{query reformulation training}, the 26,176 generated queries, including both successful and unsuccessful reformulations, facilitate the learning of adaptive search strategies. The structured traces also support \emph{student-teacher distillation}, where smaller models learn to approximate the reasoning patterns of larger analysts.

A distinguishing feature of ATC is \emph{verifiability}: every reasoning step traces to specific documents in the source corpus, enabling researchers to audit agent behavior and identify failure modes. We verify this on 5,320 trace entries. Of 1,385 substantive answers (after excluding 3,907 correct refusals where models identified insufficient evidence), 100\% are grounded in their cited documents, with a mean token coverage of 0.872, where \emph{token coverage} is the fraction of content tokens (excluding stopwords) in the synthesized answer that also appear within the cited evidence spans. The combined quality rate, grounded answers plus correct refusals, reaches 99.5\%. A finer-grained NLI-based analysis confirms 83.3\% of 6,125 atomic claims are fully supported by cited evidence. To our knowledge, no existing reasoning-trace dataset reports token-level grounding verification against source documents; prior work such as PRM800K~\cite{Lightman:2023:arXiv} verifies step correctness but not document grounding.

To validate downstream utility, we conducted student-teacher distillation using GPT-4o-generated ATC traces as the teacher signal. We fine-tuned Qwen~2.5 (0.5B to 7B) and Gemma-3 (270M) on ATC-derived supervised pairs using LoRA across 12 configurations. The smallest Qwen model (0.5B) improved its abstain-detection F1 from 0.362 to 0.815, matching the performance of a 3$\times$ larger base model. Across all sizes and architectures, trained models learn to abstain when evidence is insufficient (recall $>$ 0.88), demonstrating that the corpus captures transferable reasoning behavior.

\section{Evaluation and Results}
We address two research questions: (1) Does Corpus-Aware Seeding improve coverage compared to alternatives? (2) Can the platform reveal behavioral patterns across state-of-the-art models? We compare our seeding approach against three baselines, then analyze behavior across exploration, reformulation, and synthesis.

\subsection{Experimental Setup}
We conducted a controlled experiment using 1,000 seed queries each from three corpora: MS~MARCO (web passages about general knowledge), Quasar-T (factual questions from trivia sources), and CausalQA (questions requiring causal reasoning). Seeds were selected via our Corpus-Aware Seeding policy (Section~\ref{seeding}). We executed three simulation runs per dataset (9 total), holding the retrieval backend (ChatNoir~\cite{Bevendorff:2018:ECIR} with MS~MARCO v2.1 and ClueWeb12-22) and workflow configuration constant.

The single independent variable was the Analyst model: we rotated \texttt{\small gpt-4o}, \texttt{\small mistral-large}, and \texttt{\small deepseek-v3} through this role, with the remaining two serving as Critics in each run. For each seed query, the agent performed up to 7 retrieval-reasoning cycles of query formulation, document retrieval, relevance assessment, and optional reformulation. Each of the 3,000 seed queries produced 7 explorations via the Active Validation loop, generating a corpus of 21,000 trajectories. This design isolates behavioral differences tied to the Analyst model. Generating the full ATC consumed approximately 20M API tokens across OpenAI, Mistral, and DeepSeek endpoints, at an estimated total cost of \$200--300 USD. Individual simulations average 58 steps and \$0.14 in API costs, making corpus extension straightforward; no local GPU resources were required.

\begin{figure*}[t]
    \centering
    \includegraphics[width=\linewidth]{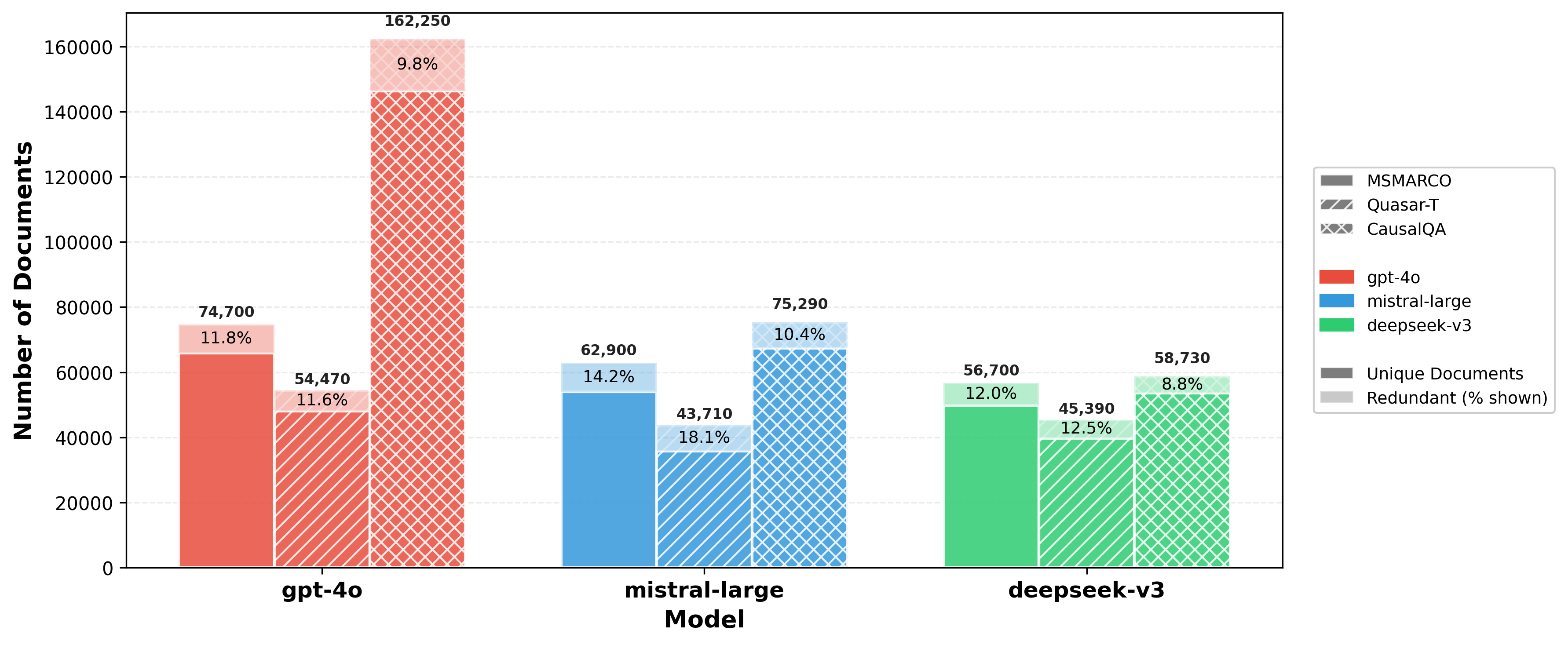}
    % \vspace{-6mm}
    \caption{Exploration breadth and retrieval redundancy across three datasets (1,000 seed queries each). Bar height indicates total retrieved documents; solid and faded regions represent unique and redundant retrievals, respectively. Percentages denote \textit{Retrieval Redundancy}, the fraction of retrievals revisiting previously seen documents.}
    \label{fig:multi-dataset-exploration}
    % \vspace{-0.1em} 
\end{figure*}

\subsection{Seeding Policy Evaluation}
We compare Corpus-Aware Seeding against three alternatives: Random (uniform sampling), Stratified (K-Means clustering with round-robin selection), and DPP (determinantal point process maximizing embedding-space diversity). Each method was evaluated across three datasets at budget $B = 500$, with five independent runs per configuration (60 total). Table~\ref{tab:seeding_baseline} reports four metrics: \emph{cluster coverage} (fraction of $K$ topic clusters whose centroid is closest to at least one selected seed), \emph{document redundancy} (mean Jaccard overlap of top-100 retrievals between selected seed pairs), \emph{semantic diversity} (mean pairwise cosine distance among seed embeddings), and \emph{corpus coverage@100} ($|\bigcup_{q\in S} \text{Top-100}(q)|/|C|$, the fraction of corpus documents reached by the top-100 retrievals of any selected seed).

Corpus-Aware achieves perfect cluster coverage (1.000, zero variance) on every dataset, while Random covers only 95--97\% and DPP drops to 45\% on CausalQA at lower budgets. All pairwise comparisons involving Corpus-Aware are statistically significant after correcting for multiple comparisons (Holm-Bonferroni), with medium-to-large effect sizes (Cohen's $d$ = 0.61--1.60). DPP achieves the highest semantic diversity, particularly on CausalQA (0.948 vs.\ 0.835 for Corpus-Aware at $B = 100$), but this advantage does not transfer to cluster or corpus coverage, confirming that optimizing for embedding-space diversity alone is insufficient for corpus construction. Stratified sampling offers minimal benefit over Random on redundancy and coverage metrics ($p > 0.14$, not significant), indicating that cluster-based allocation without novelty filtering and MMR is not enough. When matched against existing ATC traces, Corpus-Aware seeds achieve 68--98\% trace coverage compared to 2--13\% for baselines, confirming that the policy selects seeds that produce valid, non-degenerate simulations.

\subsection{Comparative Exploration Behavior}
We measured exploration behavior across all three datasets (Figure~\ref{fig:multi-dataset-exploration}), assessing two metrics: \textit{Exploration Breadth} (total unique documents retrieved) and \textit{Retrieval Redundancy} (ratio of repeated to total retrievals).

The \texttt{\small gpt-4o} Analyst demonstrated the widest exploration breadth, retrieving 31.7\% more unique documents than \texttt{\small deepseek-v3} on MS~MARCO (74,700 vs.\ 56,700), 20.0\% more on Quasar-T, and 176.3\% more on CausalQA (162,250 vs.\ 58,730). Six of nine pairwise comparisons (Mann-Whitney $U$, corrected for multiple comparisons) are statistically significant, with effect sizes up to Cohen's $d = 1.59$ (gpt-4o vs.\ deepseek-v3 on CausalQA). This gap is partly explained by differences in validation pipeline engagement: \texttt{\small deepseek-v3} triggers adaptive mode in only 0.1\% of steps, bypassing the multi-model consultation that generates alternative queries for \texttt{\small gpt-4o} (65.1\%) and \texttt{\small mistral-large} (64.0\%).

\begin{table}[tbp]
\caption{Seeding strategy comparison ($B = 500$, 5 runs). Cov.\ = cluster coverage, Red.\ = document redundancy, Div.\ = semantic diversity, CCov.\ = corpus coverage@100. All comparisons involving Corpus-Aware are significant ($p < 0.001$, Holm-Bonferroni corrected; Cohen's $d$ = 0.61--1.60).}
\label{tab:seeding_baseline}
\centering
\small
\setlength{\tabcolsep}{2.2pt}
\renewcommand{\arraystretch}{1.02}

\begin{tabularx}{\linewidth}{@{}c l *{4}{>{\centering\arraybackslash}X}@{}}
\toprule
& \textbf{Method} & \textbf{Cov.\,$\uparrow$} & \textbf{Red.\,$\downarrow$}
  & \textbf{Div.\,$\uparrow$} & \textbf{CCov.\,$\uparrow$} \\
\midrule

\multirow{4}{*}{\rotatebox[origin=c]{90}{\textit{MS MARCO}}}
& Random       & .986 & .002 & .982 & .048 \\
& Stratified   & .994 & .002 & .981 & .048 \\
& DPP          & .992 & .001 & \textbf{.987} & .047 \\
& Corpus-Aware & \textbf{1.00} & \textbf{.000} & \textbf{.987} & \textbf{.050} \\
\cmidrule(lr){2-6}

\multirow{4}{*}{\rotatebox[origin=c]{90}{\textit{Quasar-T}}}
& Random       & .968 & .007 & .971 & .026 \\
& Stratified   & .992 & .006 & .972 & .026 \\
& DPP          & .994 & .004 & \textbf{.987} & .030 \\
& Corpus-Aware & \textbf{1.00} & \textbf{.000} & .984 & \textbf{.050} \\
\cmidrule(lr){2-6}

\multirow{4}{*}{\rotatebox[origin=c]{90}{\textit{CausalQA}}}
& Random       & .968 & .095 & .785 & .044 \\
& Stratified   & .992 & .096 & .802 & .044 \\
& DPP          & .936 & .008 & \textbf{.859} & .049 \\
& Corpus-Aware & \textbf{1.00} & \textbf{.003} & .821 & \textbf{.050} \\
\bottomrule
\end{tabularx}
\end{table}

Regarding \textit{Retrieval Redundancy}, \texttt{mistral-large} was the highest (14.2\% on MS~MARCO, 18.1\% on Quasar-T, 10.4\% on CausalQA), implying a policy that favors re-validating documents. In contrast, \texttt{\small gpt-4o} maintained the lowest redundancy (11.8\%, 11.6\%, and 9.8\% respectively), which aligns with its aggressive exploration. \texttt{\small deepseek-v3} had the lowest redundancy on CausalQA (8.8\%), though this was accompanied by its reduced exploration breadth.

\subsection{Analysis of Model Reasoning Differences}
We analyzed 1,254 follow-up queries generated by \texttt{\small gpt-4o}, \texttt{\small mistral-large}, and \texttt{\small deepseek-v3}. These reformulated queries are important as they reveal each model's innate strategy for information seeking when deciding what information to seek next.

\subsubsection{Query Reformulation Strategies}
We grouped each query reformulation into three types: \textit{conceptual} (adding new related terms), \textit{procedural} (notes about the search process), and \textit{syntactic} (taking keywords or making the query simpler). Table~\ref{tab:reformulation} shows how each model behaves. The distribution of reformulation strategies differs significantly across models on all three datasets (chi-squared test, $p < 0.001$; Cram\'{e}r's $V$ = 0.054--0.073).

\begin{table}[tbp]
\caption{Distribution of query reformulation strategies.}
\label{tab:reformulation}
\centering
\small
\setlength{\tabcolsep}{3pt}
\renewcommand{\arraystretch}{1.05}

\begin{tabularx}{\linewidth}{@{}p{0.42\linewidth}*{3}{>{\centering\arraybackslash}X}@{}}
\toprule
\textbf{Model} & \textbf{Conceptual} & \textbf{Procedural} & \textbf{Syntactic} \\
\midrule
\texttt{\small mistral-large}  & \textbf{93.2\%} & \textbf{6.3\%} & 0.5\% \\
\texttt{\small gpt-4o}         & 87.1\% & 4.8\% & 8.1\% \\
\texttt{\small deepseek-v3}    & 72.4\% & 4.3\% & \textbf{23.3\%} \\
\bottomrule
\end{tabularx}
% \vspace{-2em}
\end{table}

While all models favor conceptual reformulation, \texttt{deepseek-v3} shows a high rate of syntactic simplification (23.3\%), often removing natural language structure to create keyword phrases. For example, given the query ``what was the immediate impact of the success of the manhattan project?'', \texttt{\small gpt-4o} reformulates to ``how did the manhattan project influence world politics'' (conceptual expansion), while \texttt{\small deepseek-v3} reformulates to ``Manhattan Project key scientists'' (syntactic reduction to keywords).

\subsubsection{Impact on Exploration}
This strategic difference is reflected in query length. \texttt{mistral-large} and \texttt{\small gpt-4o} expand average query length by 27\% (from 6.8 to 8.2 words) and 11\% (from 5.8 to 6.4 words) respectively, while \texttt{\small deepseek-v3} is the only model to consistently shorten queries, from 5.2 to 3.9 words ($-$16\%).

This syntactic simplification may contribute to reduced semantic exploration. Shorter, keyword-based queries may match narrower lexical spaces in the retrieval index. This aligns with our finding that \texttt{deepseek-v3} retrieved 64\% fewer unique documents than \texttt{\small gpt-4o} (58,730 vs 162,250) on the CausalQA dataset, despite generating the same number of queries. The model's reformulation strategy directly impacts the diversity of the document sets it explores.

\subsubsection{Synthesis Under Uncertainty}
Models show divergent strategies when tasked with synthesizing answers from conflicting evidence. Our analysis of 87 contradictory retrieval events across all runs reveals two distinct approaches. Models such as \texttt{\small gpt-4o} and \texttt{\small mistral-large} employ a \textit{Majority-Driven Synthesis}. They typically identify the most common claim and present it as the decisive answer. This \emph{confidence-first} approach prioritizes a clear response but obscures the underlying conflict. In contrast, \texttt{\small deepseek-v3} uses an \textit{Uncertainty-Aware Synthesis}, frequently acknowledging contradictions explicitly in its reasoning. This \emph{caution-first} stance aligns with its tendency to retrieve more documents, suggesting a more conservative policy. These are not isolated behaviors but reflect systematic differences in how models are trained to balance decisiveness against caution when faced with uncertain information.

\section{Discussion}
We examine the strengths and limitations of the AgentSim approach, and position it relative to other methods for generating agent training data.

\textbf{Strengths.}
The platform offers several advantages for RAG agent research. First, \emph{grounded verification} sets AgentSim apart from pure reasoning-trace datasets: every step can be checked against specific documents in the source corpus, allowing systematic auditing of agent behavior. Second, the Active Validation loop provides \emph{cost-effective quality control} by focusing human annotation on ambiguous cases rather than applying uniform review. The Divergence Score effectively identifies steps needing human judgment: 79.3\% of flagged seeds showed improved grounding after iterative review. Third, the modular architecture and configuration-driven execution support \emph{reproducible research}: experiments can be replicated by sharing configuration files and seed sets.

\textbf{Limitations.}
Several constraints should be noted. The approach inherits biases from its source corpora; traces generated over MS~MARCO will reflect the characteristics of web passage retrieval, which may not transfer to specialized domains. The behavioral findings we report are specific to the models evaluated; as foundation models evolve, these patterns may change. While the Active Validation loop reduces annotation burden, standardizing human review guidelines for agent reasoning remains an open challenge.

\textbf{Positioning.}
AgentSim complements, rather than competes with, existing resources. World-simulation platforms like WebArena~\cite{Zhou:2024:arXiv} capture interface-level interactions; AgentSim provides cognitive-level traces. Static benchmarks provide ground-truth labels; AgentSim transforms these into process-level supervision tailored to specific corpora, tasks, and model configurations.

\section{Conclusion}
This paper introduces AgentSim, an open-source platform for generating verifiable reasoning traces for retrieval-augmented generation (RAG) agents.
The system addresses the lack of training data for intermediate information seeking through two mechanisms: Corpus-Aware Seeding ensures systematic document exploration, while Active Validation directs human review to ambiguous instances identified by multi-model disagreement.

We release three artifacts to the research community: the AgentSim platform with its visual workflow designer, a command-line toolkit for scalable trace generation, and the Agent-Trace Corpus with over 103,000 grounded reasoning steps across three established benchmarks~\textsuperscript{\ref{fn:toolkit}}. Our evaluation shows the platform's utility for behavioral analysis, revealing systematic differences in how state-of-the-art models approach exploration, query reformulation, and synthesis under uncertainty.

Several directions merit future work. Our distillation experiments show that ATC traces improve sample efficiency across model families; scaling to pre-training is a natural next step. Domain-specific deployments for legal or scientific literature review need careful corpus curation and may reveal new challenges in agent behavior. Extending the Judge module with access to hidden solutions would let us train agents to reconstruct reasoning processes without seeing final answers. Researchers can also use ATC to benchmark their own models against our three-analyst baseline, train process reward models on per-step divergence scores, or generate domain-specific traces with the toolkit over their own corpora.

The platform, corpus, and all evaluation code are released under the MIT license\textsuperscript{\ref{fn:toolkit}}; the underlying datasets retain their original licenses. We invite the community to use AgentSim for analyzing agent behavior and generating training data, and to contribute traces back to the growing corpus.

% \begin{acks}

% \noindent
% This work has received funding from the European Union's Horizon Europe research and innovation program under grant agreement No. 101070014 (OpenWebSearch.EU), and the Bavarian State Ministry of Economic Affairs, Regional Development, and Energy (StMWi).

% % \vspace{0.5em}
% % \noindent\centering
% % \includegraphics[width=0.60\linewidth]{images/ack_hz.png}

% \end{acks}

% ---------- Bibliography ----------
\bibliographystyle{ACM-Reference-Format}
\bibliography{sample-base}

\end{document}